%% file: main.tex
\documentclass{article}
\usepackage{spconf}

\usepackage{cite}
\usepackage{amsmath,amssymb,amsfonts}
\usepackage{graphicx}
\usepackage{xcolor}
\usepackage{booktabs,array}
\usepackage{placeins}
\usepackage{dblfloatfix}
\usepackage{url}
\usepackage[hidelinks]{hyperref}
\hypersetup{
 pdftitle={ASSOCIATION PROFILE CONDITIONING IN A SET-TEMPORAL TRANSFORMER FOR CROSS-SESSION INTRACORTICAL MOTOR DECODING},
 pdfauthor={Xinyuan Zhang, Handong Mo, Pengfei Wen, Shuang Liang, Jichang Yang, Yan Zeng, Zhongrui Wang, Han Wang}}

\newcommand{\LN}{\operatorname{LN}}

\renewcommand{\dbltopfraction}{0.94}

\renewcommand{\dblfloatpagefraction}{0.70}
\makeatletter
\renewcommand\section{\@startsection{section}{1}{\z@}%
                       {-1.6ex \@plus -0.4ex \@minus -0.2ex}%
                       {0.9ex \@plus 0.2ex}%
                       {\normalfont\bfseries}}
\renewcommand\subsection{\@startsection{subsection}{2}{\z@}%
                          {-1.3ex \@plus -0.3ex \@minus -0.2ex}%
                          {0.6ex \@plus 0.2ex}%
                          {\normalfont\bfseries}}
\renewcommand\subsubsection{\@startsection{subsubsection}{3}{\z@}%
                             {-1.0ex \@plus -0.2ex \@minus -0.1ex}%
                             {0.4ex \@plus 0.1ex}%
                             {\normalfont\bfseries}}
\newcommand{\APSTIntroFigureRightColumn}{%
  \if@firstcolumn\suppressfloats[t]\fi
}
\makeatother
\makeatletter
\long\def\@makecaption#1#2{%
  \vskip\abovecaptionskip
  \setbox\@tempboxa\hbox{#1. #2}%
  \ifdim\wd\@tempboxa>\hsize
    #1. #2\par
  \else
    \hbox to\hsize{\hfil\box\@tempboxa\hfil}%
  \fi
  \vskip\belowcaptionskip
}
\makeatother
\makeatletter
\renewcommand{\title}[1]{\gdef\@title{#1}}
\renewcommand{\name}[1]{\gdef\@name{#1}}
\renewcommand{\@maketitle}{%
  \vspace*{0.6em}%
  \begin{center}
    {\fontsize{13.0}{14.5}\selectfont\bfseries \@title\par}%
    \vspace{0.45em}%
    {\fontsize{9.5}{10.5}\selectfont\itshape \@name\par}%
    \vspace{0.2em}%
    {\fontsize{9}{10.5}\selectfont \@address\par}%
  \end{center}%
  \vspace{0.35em}%
}
\makeatother

\begin{document}
\ninept
\title{ASSOCIATION PROFILE CONDITIONING IN A SET-TEMPORAL TRANSFORMER\\FOR CROSS-SESSION INTRACORTICAL MOTOR DECODING}
\name{Xinyuan Zhang$^{1,*}$, Handong Mo$^{2,*}$, Pengfei Wen$^{2,*}$, Shuang Liang$^{1}$, Jichang Yang$^{1}$, Yan Zeng$^{1}$, Zhongrui Wang$^{2,\dagger}$, Han Wang$^{1,\dagger}$}
\address{$^{1}$The University of Hong Kong, Hong Kong SAR, China\quad
$^{2}$Southern University of Science and Technology, Shenzhen, China\\
$^{*}$Equal contribution.\quad
$^{\dagger}$Correspondence: \texttt{wangzr@sustech.edu.cn}; \texttt{hanwang6@hku.hk}}
\maketitle

\begin{abstract}
\input{manuscript/abstract}
\end{abstract}

\begingroup
\emergencystretch=3em
\input{manuscript/01_introduction}
\par
\endgroup

\begin{figure*}[t]
\centering
\includegraphics[width=0.92\textwidth]{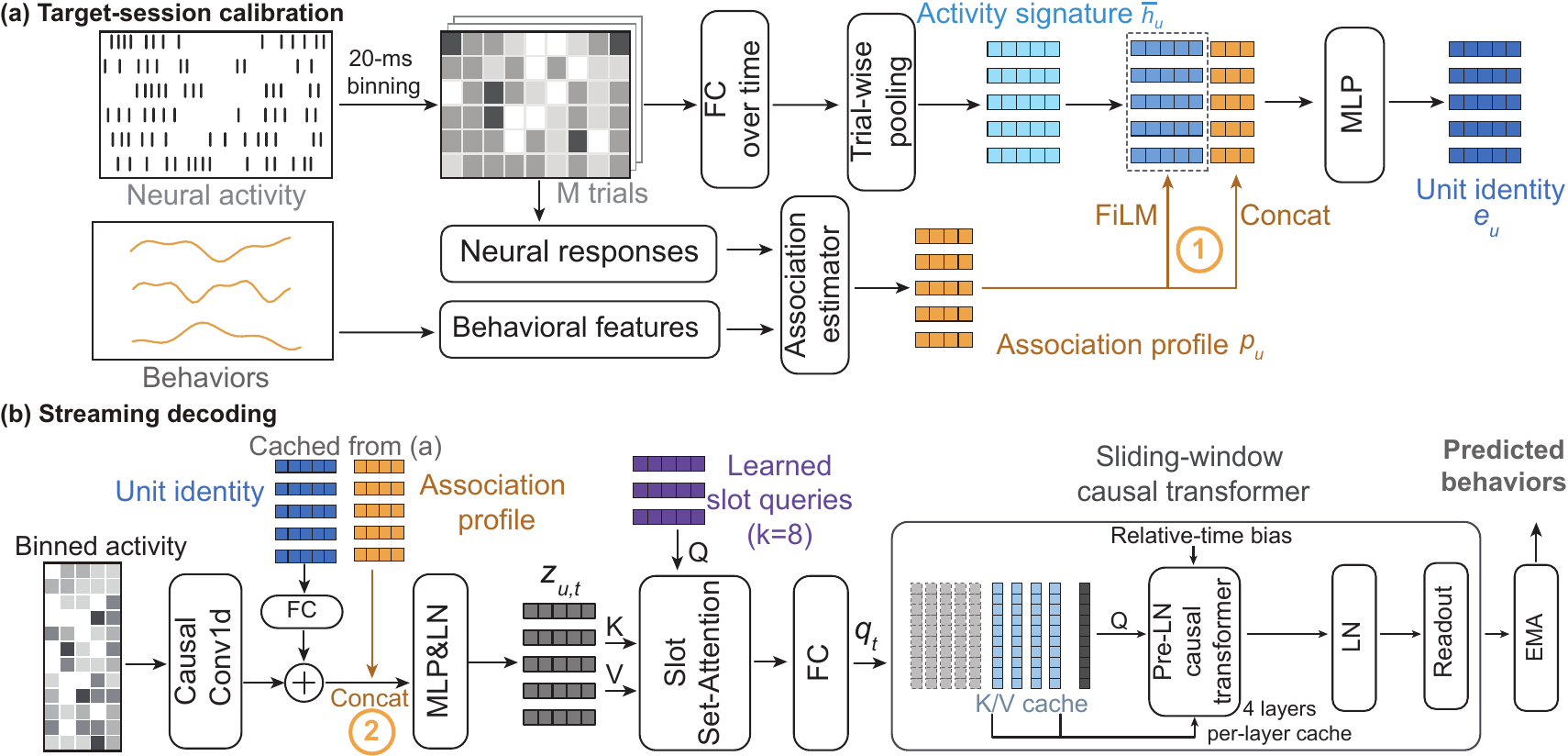}
\caption{APST overview. (a) Target-session calibration: from $M$ labeled trials, the association estimator produces each unit's profile $p_u$; the profile conditions the activity signature at site \textcircled{\scriptsize 1} through FiLM and concatenation, forming unit identity $e_u$. (b) Streaming decoding: live activity is combined with the cached identity and profile at site \textcircled{\scriptsize 2}, aggregated across units by set attention with learned slot queries, then decoded by a sliding-window causal transformer.}
\label{fig:network}
\end{figure*}

\input{manuscript/02_methods}

\begin{figure*}[!t]
\centering
\includegraphics[width=\textwidth]{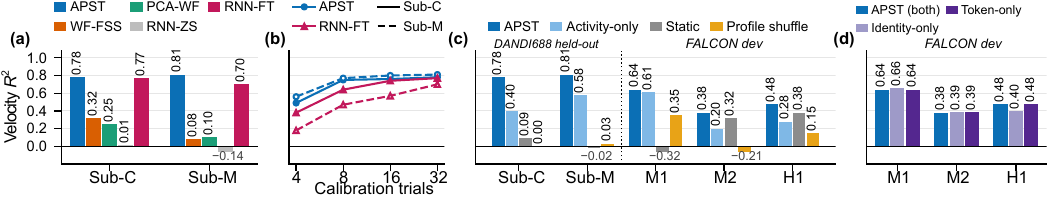}
\caption{Results on DANDI688 and FALCON (shared y-axis: velocity $R^2$). (a) DANDI688 held-out sessions at 32 calibration trials. (b) Calibration curves (solid: Sub-C, dashed: Sub-M; blue: APST, pink: RNN-FT). (c) Ablations on DANDI688 held-out and FALCON dev. (d) Conditioning-site ablation on FALCON dev. (c,d) Three-seed means; negative bars are truncated with values annotated.}
\label{fig:sua_comparison}
\end{figure*}

\input{manuscript/03_experiments}
\input{manuscript/04_discussion}

\FloatBarrier
\clearpage
\let\oldthebibliography\thebibliography
\let\endoldthebibliography\endthebibliography
\renewenvironment{thebibliography}[1]{%
  \begin{oldthebibliography}{#1}%
    \fontsize{8.2pt}{9.4pt}\selectfont
    \setlength{\itemsep}{0pt plus 0.2pt}%
    \setlength{\parsep}{0pt}%
    \setlength{\parskip}{0pt}%
}{%
  \end{oldthebibliography}%
}

\bibliographystyle{IEEEbib}
\bibliography{references}

\section*{Acknowledgments}
This work was supported by the Research Grants Council of Hong Kong under grants AoE/E-101/23-N, C1009-22G, and T45-701/22-R. The authors have no relevant financial or nonfinancial interests to disclose.

\section*{Compliance with Ethical Standards}
This study used only publicly released recordings and conducted no new human or animal experiments. The data are the FALCON M1, M2, and H1 tasks~\cite{falcon2024,rouse2016m1,nason2021fingerbmi,collinger2013} and DANDI Archive dataset 000688~\cite{gallego2020,perich2025dandi}. Ethical approval for the original recordings was obtained by the data collectors.
\end{document}

%% file: manuscript/abstract.tex
Intracortical motor decoders degrade across sessions because the set of recorded units changes and persisting units can alter how their firing relates to behavior. Most existing methods update network weights on each new session or rely on unlabeled activity, which does not directly reveal such changes. We present APST, an \textbf{A}ssociation \textbf{P}rofile-conditioned \textbf{S}et-\textbf{T}emporal transformer that adapts to new sessions with all network weights frozen. From a few labeled calibration trials, APST summarizes how each unit's firing relates to behavior in a four-dimensional association profile computed in closed form. The profiles condition a set-attention encoder that accepts any number and order of units, followed by a causal transformer for streaming decoding. On held-out DANDI688 sessions from two monkeys, APST reaches velocity $R^2$ of $0.78$ and $0.81$, versus $0.40$ and $0.58$ for a variant that uses neural activity alone, and matches or exceeds an RNN fine-tuned on the same trials. On FALCON private held-out evaluation, it attains $R^2$ of $0.65$, $0.42$, and $0.44$ on M1, M2, and H1.

%% file: manuscript/01_introduction.tex
\section{Introduction}
\begin{figure}[t]
\centering
\includegraphics[width=\columnwidth]{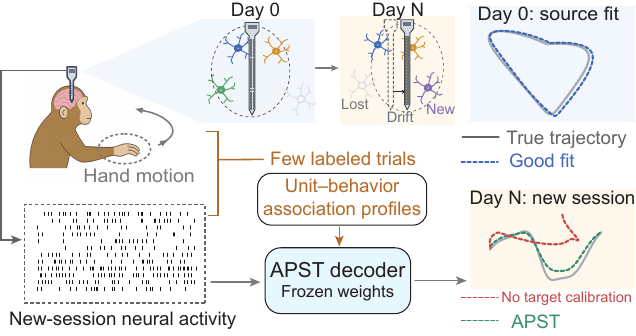}
\caption{Across sessions, electrode drift changes the recorded units, and a
decoder fit on Day~0 degrades on Day~$N$ without target calibration. APST
adapts with frozen weights by conditioning on association profiles estimated
from a few labeled calibration trials.}
\label{fig:intro}
\end{figure}

Intracortical motor decoding maps neural activity to continuous motor output, such as limb kinematics or muscle activity. Slight array motion relative to tissue~\cite{steinmetz2021,yuan2024} changes the set of recorded units (neurons or channels) across sessions (recording days). Units are lost, gained, or drift, and persisting units can change their associations with behavior~\cite{falcon2024} (Fig.~\ref{fig:intro}). Although the low-dimensional motor manifold is comparatively stable~\cite{gallego2020}, a decoder fit on Day~0 degrades on Day~$N$ without target-session calibration. EEG adaptation often assumes a fixed channel layout~\cite{jiang2024eeg,rahmani2026eeg}; in intracortical recordings, a fixed electrode index need not denote the same neural population across sessions.

\APSTIntroFigureRightColumn

Existing approaches adapt pretrained decoders through latent alignment~\cite{degenhart2020,farshchian2019,nomad2025} or target-session fine-tuning~\cite{ye2023ndt2}. Permutation-invariant architectures accept unit sets that change across sessions~\cite{poyo2023}, and SPINT infers unit embeddings from unlabeled calibration activity without target-session gradient updates~\cite{spint2025}. Unlabeled activity, however, cannot resolve all drift: a persisting unit may keep its firing statistics while its association with behavior changes~\cite{rokni2007}. Behavioral labels can resolve it, and they are routinely available: clinical intracortical BCIs are periodically recalibrated with short blocks of cued movements~\cite{jarosiewicz2015}, and FALCON~\cite{falcon2024} formalizes this setting by releasing a few labeled calibration trials at the start of each held-out session. Labeled few-shot fine-tuning~\cite{ye2023ndt2} exploits these labels, but only by updating network weights on each new session, which requires backpropagation at deployment and risks overfitting a few trials. Whether a few labeled trials can adapt a decoder whose weights stay frozen remains unclear.

Each unit's association with behavior can be summarized from a few labeled calibration trials. Directional tuning relates a unit's firing rate to movement direction and yields a preferred direction and modulation depth~\cite{georgopoulos1982}. Encoding models relate neural responses to behavioral covariates more generally~\cite{truccolo2005,hatsopoulos2007}; for multidimensional kinematics or muscle EMG, behavior-weighted associations are projected onto a source-fitted singular value decomposition (SVD) basis. Either way, each unit receives an association profile, a task-conditioned functional identity that complements the activity signatures estimated from the same trials.

We introduce APST\footnote{{\fontsize{6.5}{8}\selectfont\mbox{Source code is available at \url{https://github.com/HsinyuanZhang/APST}.}}}, an \textbf{A}ssociation \textbf{P}rofile-conditioned \textbf{S}et-\textbf{T}emporal transformer for cross-session decoding without target-session backpropagation. Specifically: \textbf{(i)} APST estimates a 4-D association profile for each unit from a few labeled calibration trials in closed form, with all source-trained weights frozen. \textbf{(ii)} The profiles condition a permutation-invariant set-attention encoder that accepts unit sets of any size, followed by a sliding-window causal temporal transformer for streaming motor decoding. \textbf{(iii)} We evaluate APST on single-unit recordings from DANDI Archive 000688 (DANDI688; Sub-C and Sub-M)~\cite{gallego2020,perich2025dandi} and official FALCON benchmarks (M1, M2, H1)~\cite{falcon2024}. Ablations show that association profiles raise velocity $R^2$ from $0.40$ to $0.78$ on Sub-C over an activity-only variant, and that profiles shuffled across units score below activity-only.

%% file: manuscript/02_methods.tex
\section{Method}
\label{sec:profile}

Calibration (Fig.~\ref{fig:network}(a)) gives each observed unit a static profile and a cached identity; streaming decoding (Fig.~\ref{fig:network}(b)) combines them with live activity and decodes causally.

\subsection{Notation and calibration boundary}
Let $s$ index sessions and $u=1,\ldots,N_s$ index observed units (sorted single units or unsorted channels). A target calibration set provides $M_s$ labeled trials $\mathcal C_s=\{(X_{s,j},Y_{s,j})\}_{j=1}^{M_s}$, with neural activity $X_{s,j}$ and behavioral targets $Y_{s,j}$. All network parameters are trained on source sessions and frozen during target deployment. From $\mathcal C_s$, the system computes a static 4-D \emph{association profile} $p_{s,u}$ and unit identity $e_{s,u}$ without target-session backpropagation.

\subsection{Association profiles}
An association profile summarizes how a unit's calibration activity relates to task behavior ($\text{activity} \xrightarrow{\text{behavior}} \text{coefficients} \xrightarrow{\text{map}} \text{profile}$). Directional profiles combine three fitted coefficients with their derived modulation depth, while multidimensional association summaries are projected to the same four-dimensional interface.

Encoding models relate unit responses to behavioral covariates~\cite{truccolo2005,hatsopoulos2007}. Let $r_{s,t,u}$ be the response of unit $u$ at calibration sample $t$ (a trial or time bin). The task supplies $\phi_{s,t}\in\mathbb R^{d_\phi}$ and nonnegative weight $\omega_{s,t}$. The raw coefficient vector is
\newcommand{\TightenDisplay}{%
  \setlength{\abovedisplayskip}{4pt plus 1pt minus 1pt}%
  \setlength{\belowdisplayskip}{4pt plus 1pt minus 1pt}%
  \setlength{\abovedisplayshortskip}{0pt plus 1pt}%
  \setlength{\belowdisplayshortskip}{4pt plus 1pt minus 1pt}%
}
\begingroup
\TightenDisplay
\begin{equation}\label{eq:gen-profile}
a^{\mathrm{raw}}_{s,u}=\!\underset{v\in\mathbb R^{d_\phi}}{\operatorname{argmin}}\!\sum_{t}\omega_{s,t}(r_{s,t,u}-\phi_{s,t}^{\top}v)^{2}+\eta\|v\|_2^{2}.
\end{equation}
\endgroup
The closed-form solution requires one calibration pass and no target-session gradients. Behavior enters through either the design $\phi$ or the weights $\omega$, and the two estimators below differ in this choice.

\subsection{Directional tuning profiles (M2 and DANDI688)}
For two-dimensional movements, behavior enters the design via trial movement direction $\theta_{s,j}$. With $\phi_{s,j}=[1,\cos\theta_{s,j},\sin\theta_{s,j}]$, uniform weights, and $\eta\to0$, Eq.~\eqref{eq:gen-profile} corresponds to first-harmonic least-squares tuning~\cite{georgopoulos1982}:
\begin{equation}\label{eq:fx-direction-fit}
\widehat R_{s,j,u} = b_{s,u} + a_{s,u}\cos\theta_{s,j} + d_{s,u}\sin\theta_{s,j},\; \rho_{s,u} = \sqrt{a_{s,u}^2 + d_{s,u}^2},
\end{equation}
where $R_{s,j,u}$ is the movement-window mean count per bin. The profile collects the directional coefficients, their modulation depth, and the baseline response:
\begingroup
\TightenDisplay
\begin{equation}\label{eq:fx-direction-pack}
p_{s,u} = \bigl([a_{s,u},\ d_{s,u},\ \rho_{s,u},\ b_{s,u}]^\top - \mu^{\mathrm{dir}}\bigr) \oslash \sigma^{\mathrm{dir}},
\end{equation}
\endgroup
where ${}^{\top}$ denotes transposition, $\oslash$ element-wise division, and $\mu^{\mathrm{dir}}$ and $\sigma^{\mathrm{dir}}$ the source-session mean and standard deviation.

\subsection{SVD-compressed association profiles (M1 and H1)}
For M1 (16-channel EMG) and H1 (7-D velocity), behavior enters through sample weights $\omega_{s,t}=w_{s,t,k}$ for each condition $k$, and Eq.~\eqref{eq:gen-profile} is evaluated on standardized unit responses $\tilde r$. M1 uses $w_{s,t,k}=\max(\mathrm{EMG}_{s,t,k},0)/\mathrm{RMS}^{\mathrm{src}}_k$; H1 divides velocity by its source RMS, then takes softplus positive and negative states, yielding $K=16$ for M1 and $K=14$ for H1.

We normalize unit responses, compute behavior-weighted associations, and project the resulting matrix to four dimensions:
\begingroup
\TightenDisplay
\begin{equation}\label{eq:fx-weighted-profile}
\begin{aligned}
\sigma_{s,u}^{\mathrm{rate}} &= \frac{\sqrt{\max(\Delta\bar r_{s,u},1)}}{\Delta}, \qquad \tilde r_{s,t,u} = \frac{r_{s,t,u}-\bar r_{s,u}}{\sigma_{s,u}^{\mathrm{rate}}},\\
A_{s,u,k} &= \frac{\sum_{t}w_{s,t,k}\,\tilde r_{s,t,u}}{\sum_{t}w_{s,t,k}+\eta},\\
P_s &= (A_sV_4-\mathbf1\mu^\top)\operatorname{diag}(\nu)^{-1}.
\end{aligned}
\end{equation}
Here $r_{s,t,u}$ is firing rate (in Hz) with calibration mean $\bar r_{s,u}$, and $\sigma_{s,u}^{\mathrm{rate}}$ is a Poisson-floor rate scale with effective bin width $\Delta$. In Eq.~\eqref{eq:gen-profile}, fitting each state with scalar design ($\phi\equiv1$, $\omega_{s,t}=w_{s,t,k}$) on $\tilde r$ yields association matrix $A_s\in\mathbb R^{N_s\times K}$, shrunk toward zero when $\sum_t w_{s,t,k}$ is small. Projecting onto the top four right singular vectors $V_4\in\mathbb R^{K\times 4}$ of uncentered pooled source associations $A_{\mathrm{src}}=U\Sigma V^\top$ and standardizing by source moments ($\mu,\nu$) outputs profile matrix $P_s=[p_{s,1}^\top;\ldots;p_{s,N_s}^\top]\in\mathbb R^{N_s\times 4}$, matching the 4-D interface. M1 uses one shared post-projection source RMS scale and H1 coordinatewise source standard deviations.

\input{tables/table_falcon_benchmark}

\subsection{Profile-conditioned unit identities}
Following calibration-based unit representations~\cite{spint2025}, we combine activity signatures with association profiles to form task-conditioned unit identities. Calibration activity $x_{s,j,u}\in\mathbb R^{B_{\mathrm{cal}}}$ (binned counts of unit $u$ in calibration trial $j$) passes through a shared layer and is trial-averaged into the activity signature $\overline h_{s,u}=M_s^{-1}\sum_{j=1}^{M_s}\operatorname{ReLU}(W_a x_{s,j,u}+b_a)$.

At identity-conditioning site \textcircled{\scriptsize 1} in Fig.~\ref{fig:network}(a), profile $p_{s,u}$ modulates this signature through FiLM~\cite{perez2018film} and is concatenated before the identity MLP $g$:
\begin{equation}\label{eq:fx-film-identity}
\begin{aligned}
{[\gamma_{s,u};\beta_{s,u}]}&=W_o\operatorname{ReLU}(W_i p_{s,u}+b_i)+b_o,\\
\tilde h_{s,u} &= (1+\gamma_{s,u})\odot\overline h_{s,u}+\beta_{s,u},\\
e_{s,u}&=g([\tilde h_{s,u};\ p_{s,u}]).
\end{aligned}
\end{equation}
Here $[\cdot;\cdot]$ denotes concatenation and $\odot$ element-wise multiplication. FiLM lets the profile scale and shift individual activity features, while concatenation passes it to $g$ directly. Following adaLN-Zero~\cite{peebles2023dit}, $W_o$ and $b_o$ are zero-initialized, so training starts from concatenation alone; the unit identities $e_{s,u}$ are computed once and cached.

\subsection{Set-temporal streaming decoder}
At token-conditioning site \textcircled{\scriptsize 2} in Fig.~\ref{fig:network}(b), a shared causal convolution over five bins maps live activity to $l_{s,u,t}=\psi(x_{s,u,t-4:t})$. The projected identity is added to this activity representation, and $p_{s,u}$ is concatenated before the token MLP $f$ and layer normalization $\operatorname{LN}$ form the unit token:
\begin{equation}\label{eq:fx-token}
z_{s,u,t}=\operatorname{LN}\!\left(f([l_{s,u,t}+W_e e_{s,u};p_{s,u}])\right).
\end{equation}
Eight learned slot queries $S$ aggregate the unit tokens $Z_{s,t}$ with multi-head set attention~\cite{settransformer2019,perceiver2021}. Set aggregation is invariant to jointly permuting each unit's activity, identity, and association profile. Padding mask $m_s$ excludes padded units before softmax, which permits variable $N_s$:
\begin{equation}\label{eq:fx-set}
\begin{aligned}
\widetilde S_{s,t}&=\LN(S)+\operatorname{MHA}(\LN(S),Z_{s,t},Z_{s,t};m_s),\\
q_{s,t}&=W_q\operatorname{vec}\!\left(\widetilde S_{s,t}+\operatorname{FFN}(\operatorname{LN}(\widetilde S_{s,t}))\right).
\end{aligned}
\end{equation}
\endgroup
Here $\operatorname{MHA}$ computes cross-attention from queries $S$ to unit tokens $Z_{s,t}$ under mask $m_s$, and $\operatorname{vec}$ flattens the slot matrix before linear projection $W_q$ produces the 256-D population token $q_{s,t}$.

Starting from $H^{0}_{s,t}=q_{s,t}$, four pre-LN causal transformer layers with eight heads process the population sequence through residual attention and FFN sublayers~\cite{vaswani2017}. At time $t$, layer $\ell$ attends to the key/value pairs at $t$ and the preceding $w_\ell-1$ bins. Within each causal window, attention logits include a learned relative-time bias inspired by ALiBi~\cite{press2022alibi}. These local causal windows define the raw decoder receptive field $R=5+\sum_{\ell}(w_\ell-1)$, including the five-bin convolution. The context length $R$ is set from the temporal structure of each dataset and split evenly across the four causal layers. Per-layer key/value caches bound streaming state by this receptive field rather than the full session history. During source training, whole-unit dropout removes random units from the set.

A final layer normalization and task-specific readout give the raw prediction $\widehat y_{s,t}=r_{\mathrm{task}}(\operatorname{LN}(H^{4}_{s,t}))$, which an inference-only causal EMA smooths as $\widetilde y_{s,t}=\alpha\widetilde y_{s,t-1}+(1-\alpha)\widehat y_{s,t}$. The EMA ($\alpha=1/3$) resets at session boundaries, starting from $\widetilde y_{s,0}=\widehat y_{s,0}$.

%% file: tables/table_falcon_benchmark.tex
\begin{table*}[!b]
\centering
\setlength{\abovecaptionskip}{3pt}
\setlength{\belowcaptionskip}{4pt}
\caption{Official FALCON private held-out performance. Entries are mean $\pm$ SD $R^2$ and normalized latency (lat.).}
\label{tab:falcon}
\footnotesize
\setlength{\tabcolsep}{2pt}
\renewcommand{\arraystretch}{0.85}
\begin{tabular*}{\textwidth}{@{\extracolsep{\fill}}>{\raggedright\arraybackslash}p{1.42in}>{\raggedright\arraybackslash}p{1.25in}cccccc@{}}
\toprule
Method & Target adaptation & \multicolumn{2}{c}{M1 ($R^2$ / lat.)} & \multicolumn{2}{c}{M2 ($R^2$ / lat.)} & \multicolumn{2}{c}{H1 ($R^2$ / lat.)}\\
\midrule
\multicolumn{8}{l}{\textit{Gradient-free target-session adaptation}}\\
WF & None & $0.34\pm0.06$ & 0.06 & $0.06\pm0.04$ & 0.08 & $0.16\pm0.03$ & 0.15\\
RNN & None & $-0.60\pm0.45$ & 0.03 & $-0.07\pm0.23$ & 0.01 & $0.09\pm0.18$ & 0.02\\
SPINT & Unlabeled few-shot & $0.66\pm0.07$ & 0.13 & $0.26\pm0.13$ & 0.13 & $0.29\pm0.15$ & 0.14\\
\textbf{APST (ours)} & Labeled few-shot & $\mathbf{0.65\pm0.11}$ & 0.15 & $\mathbf{0.42\pm0.10}$ & 0.09 & $\mathbf{0.44\pm0.15}$ & 0.15\\
\midrule
\multicolumn{8}{l}{\textit{Gradient-based target adaptation}}\\
CycleGAN + WF & Unlabeled few-shot & $0.43\pm0.04$ & 0.07 & $0.22\pm0.06$ & 0.09 & $0.12\pm0.06$ & 0.16\\
NoMAD + WF & Unlabeled few-shot & $0.49\pm0.03$ & 0.99 & $0.20\pm0.10$ & 0.91 & $0.13\pm0.10$ & 1.03\\
NDT2 Multi & Labeled few-shot & $0.59\pm0.07$ & 0.13 & $0.43\pm0.08$ & 0.10 & $0.52\pm0.04$ & 0.30\\
\midrule
\multicolumn{8}{l}{\textit{Private-label oracle references}}\\
WF & Private labels & $0.53\pm0.04$ & 0.06 & $0.26\pm0.03$ & 0.08 & $0.21\pm0.04$ & 0.14\\
RNN & Private labels & $0.75\pm0.05$ & 0.04 & $0.56\pm0.04$ & 0.04 & $0.44\pm0.13$ & 0.08\\
NDT2 Multi & Private labels & $0.78\pm0.04$ & 0.15 & $0.58\pm0.04$ & 0.10 & $0.63\pm0.08$ & 2.29\\
\bottomrule
\end{tabular*}
\par\smallskip
{\raggedright\footnotesize\textit{Note.} Official private held-out evaluation. Bold indicates our method. Other methods from~\cite{spint2025}.\par}
\end{table*}

%% file: manuscript/03_experiments.tex
\section{Experiments}

\subsection{Datasets and protocol}
We evaluate APST across four continuous motor-decoding benchmarks binned at 20~ms: FALCON~\cite{falcon2024} M1 (rhesus primary motor cortex to 16-channel upper-limb EMG)~\cite{rouse2016m1}, M2 (rhesus Utah arrays to 2-D finger-group velocity)~\cite{nason2021fingerbmi}, H1 (human Utah arrays to 7-D robotic arm and hand velocity)~\cite{collinger2013}, and DANDI Archive 000688~\cite{gallego2020,perich2025dandi} (rhesus Utah array to 2-D cursor velocity). Sub-C and Sub-M are two different monkeys recorded under this same configuration. FALCON inputs are unsorted threshold crossings, scored via official EvalAI means on private evaluation data from held-out sessions, with reported SD; normalized latency is processing time divided by evaluation-data duration. DANDI688 uses sorted single units.

On DANDI688, Sub-C uses 18 source, 6 development, and 6 held-out sessions; Sub-M uses 6 source, 2 development, and 3 held-out sessions. Ablations and linear baselines use 32 labeled calibration trials per target session; comparisons with the RNN (Fig.~\ref{fig:sua_comparison}(a,b), Table~\ref{tab:heldout}) use budgets of 4, 8, 16, and 32 trials. Performance is variance-weighted velocity $R^2$ per session followed by an equal-session mean. Hyperparameters are chosen per dataset on local development splits, primarily tuning the learning rate, the whole-unit dropout rate, the projection dimension of the identity mapping $W_e$ into the decoding stream, and the slope initialization coefficients of the relative-time bias.

\subsection{Main results}
Table~\ref{tab:falcon} reports official FALCON results, with baselines and oracle references from SPINT~\cite{spint2025} and FALCON~\cite{falcon2024}. With all weights frozen, APST achieves $0.65\pm0.11$, $0.42\pm0.10$, and $0.44\pm0.15$ on M1, M2, and H1. Among methods without target-session gradients, it has the highest mean on M2 and H1 and is within $0.01$ of SPINT on M1. It also exceeds the gradient-based NDT2 Multi~\cite{ye2023ndt2} on M1, comes within $0.01$ of it on M2, and matches the private-label RNN oracle on H1. Unlike SPINT, APST uses target-session labels; NDT2 Multi uses the same labeled calibration trials but updates network weights, whereas APST uses them only through closed-form profiles. Normalized latencies range from $0.09$ to $0.15$.

Table~\ref{tab:falcon-held-in} reports official private held-in means (other entries from SPINT Table~A1~\cite{spint2025}). Because held-in sessions contribute source-training data, the held-in minus held-out gap ($\mathrm{HI}-\mathrm{HO}$) measures the accuracy lost on unseen sessions. APST's held-in $R^2$ ($0.77, 0.64, 0.60$) matches the strongest baselines (tied best on M1, best on M2, second to NDT2 Multi on H1), so its held-out gains in Table~\ref{tab:falcon} do not sacrifice within-session accuracy. Against SPINT, the other gradient-free method with competitive held-in accuracy, APST loses less on M2 ($0.22$ vs.\ $0.33$) and H1 ($0.16$ vs.\ $0.18$) and a similar amount on M1 ($0.12$ vs.\ $0.11$). The smaller gaps of WF and CycleGAN + WF reflect their low held-in accuracy rather than robustness.

\input{tables/table_falcon_held_in}

\looseness=-1
On DANDI688 held-out sessions (Fig.~\ref{fig:sua_comparison}(a)), we compare APST with linear baselines, WF-FSS (Wiener filter with full session support) and PCA-WF (principal component analysis Wiener filter)~\cite{glaser2020,cunningham2014,falcon2024}, and an RNN baseline using FALCON's architecture and search grid~\cite{falcon2024}. Zero-shot RNN (RNN-ZS) yields negligible accuracy ($0.01$ on Sub-C, $-0.14$ on Sub-M) due to unaligned unit channels, but rises to $0.77$ and $0.70$ when fine-tuned on target calibration trials (RNN-FT). With frozen weights, APST reaches $0.78$ on Sub-C and $0.81$ on Sub-M, matching or exceeding RNN-FT and outperforming the linear baselines in Fig.~\ref{fig:sua_comparison}(a).

\begin{table}[t]
\centering
\setlength{\abovecaptionskip}{3pt}
\setlength{\belowcaptionskip}{4pt}
\caption{Held-out velocity $R^2$ at 32 calibration trials. Act-only: Activity-only (no target labels); Dates in 2015 (same sessions as Fig.~\ref{fig:sua_comparison}, different seeds).}
\label{tab:heldout}
\scriptsize
\setlength{\tabcolsep}{1.6pt}
\renewcommand{\arraystretch}{0.80}
\begin{tabular*}{\columnwidth}{@{\extracolsep{\fill}}lcccccc@{}}
\toprule
Sub-C & 11-13 & 11-16 & 11-17 & 11-19 & 11-20 & 12-01 \\
\textbf{APST} & \textbf{0.80} & \textbf{0.78} & \textbf{0.82} & \textbf{0.87} & \textbf{0.83} & \textbf{0.58} \\
Act-only & 0.56 & 0.58 & 0.65 & $-0.02$ & 0.46 & 0.06 \\
RNN-FT & 0.83 & 0.81 & 0.83 & 0.83 & 0.85 & 0.48 \\
\midrule
Sub-M & 06-23 & 06-25 & 06-26 & & & \\
\textbf{APST} & \textbf{0.84} & \textbf{0.79} & \textbf{0.80} & & & \\
Act-only & 0.49 & 0.60 & 0.66 & & & \\
RNN-FT & 0.73 & 0.69 & 0.69 & & & \\
\bottomrule
\end{tabular*}
\end{table}

\looseness=-1
Sub-M's held-out sessions come 8 to 11 days after its last source session, and APST outscores Activity-only and RNN-FT on all three (Table~\ref{tab:heldout}). Sub-C's come 120 to 138 days after, and there RNN-FT is higher on four of six sessions, by at most $0.03$. On the most distant session (12-01), where every method declines, APST retains $0.58$ versus $0.48$ for RNN-FT and $0.06$ for Activity-only.

\looseness=-1
On the calibration-budget curves (Fig.~\ref{fig:sua_comparison}(b)), we compare APST with RNN fine-tuning across $4$--$32$ labeled trials. At 8 trials, APST reaches $0.75$ on Sub-C and $0.77$ on Sub-M, versus $0.64$ and $0.47$ for RNN-FT. On Sub-C, 8 APST trials exceed the $0.74$ that fine-tuning reaches with 16 trials. On Sub-M, APST stays ahead at every budget by $0.11$ to $0.38$, indicating greater sample efficiency.

\looseness=-1
Target-session adaptation updates no network weights. With unit sets padded to $100$ for both subjects, calibration costs $3.83$ to $21.75$ million multiply--accumulate operations (MACs) for 4 to 32 trials, of which the closed-form profile fit accounts for $<0.1\%$ ($<15$k operations). Fine-tuning the RNN updates all $118$k weights on Sub-C and $367$k on Sub-M through 512 updates of batch size 128 over 50-bin windows, totaling $383$ and $1195$ billion forward MACs.

\subsection{Ablation studies}
\looseness=-1
Fig.~\ref{fig:sua_comparison}(c,d) evaluates ablations on DANDI688 held-out sessions and FALCON development splits. \emph{Activity-only} is trained and calibrated without profiles, conditioning units on activity signatures alone. \emph{Static} uses one source-trained identity table in place of calibrated identities. \emph{Profile shuffle} reassigns profiles across units.

\looseness=-1
In Fig.~\ref{fig:sua_comparison}(c), profiles raise performance over Activity-only from $0.40$ to $0.78$ on Sub-C and from $0.58$ to $0.81$ on Sub-M, and add $0.03$, $0.18$, and $0.20$ on M1, M2, and H1. Shuffled profiles score below Activity-only in all five settings, reaching $0.00$ and $0.03$ on Sub-C and Sub-M. Because shuffling leaves each session's set of profiles unchanged, this drop shows that APST uses each profile as a unit-specific identity rather than as a session-level summary. Static ranges from $-0.32$ to $0.38$ across datasets, and APST exceeds it in every setting, with the smallest margins of $0.06$ on M2 and $0.10$ on H1. On DANDI688, units are sorted anew in each session, so a single identity table cannot follow them across sessions.

\looseness=-1
Fig.~\ref{fig:sua_comparison}(d) ablates the conditioning site on FALCON over three seeds. Identity-only applies $p_{s,u}$ only at site~\textcircled{\scriptsize 1}, Token-only only at site~\textcircled{\scriptsize 2}, and APST at both. Token-only and dual conditioning differ by at most $0.01$ on every task, within the seed SD of $0.02$, whereas Identity-only falls to $0.40$ on H1 against $0.48$; token conditioning therefore carries most of the benefit. We retain site~\textcircled{\scriptsize 1} because it adds no streaming cost: the identity is computed once during calibration and cached, and the zero-initialized FiLM starts as an identity map. All reported results use dual conditioning.

%% file: tables/table_falcon_held_in.tex
\begin{table}[t]
\centering
\setlength{\abovecaptionskip}{3pt}
\setlength{\belowcaptionskip}{4pt}
\caption{Official private held-in mean $R^2$. Parentheses denote held-in minus held-out gap ($\mathrm{HI}-\mathrm{HO}$). Bold indicates our method.}
\label{tab:falcon-held-in}
\scriptsize
\setlength{\tabcolsep}{2.5pt}
\renewcommand{\arraystretch}{0.80}
\begin{tabular}{@{}lccc@{}}
\toprule
Method & M1 & M2 & H1 \\
\midrule
WF & $0.46$ ($0.12$) & $0.15$ ($0.09$) & $0.20$ ($0.04$) \\
RNN & $0.52$ ($1.12$) & $0.20$ ($0.27$) & $0.31$ ($0.22$) \\
SPINT & $0.77$ ($0.11$) & $0.59$ ($0.33$) & $0.47$ ($0.18$) \\
CycleGAN + WF & $0.61$ ($0.18$) & $0.32$ ($0.10$) & $0.15$ ($0.03$) \\
NoMAD + WF & $0.64$ ($0.15$) & $0.35$ ($0.15$) & $0.21$ ($0.08$) \\
NDT2 Multi & $0.77$ ($0.18$) & $0.63$ ($0.20$) & $0.62$ ($0.10$) \\
\midrule
\textbf{APST (ours)} & $\mathbf{0.77}$ ($\mathbf{0.12}$) & $\mathbf{0.64}$ ($\mathbf{0.22}$) & $\mathbf{0.60}$ ($\mathbf{0.16}$) \\
\bottomrule
\end{tabular}
\end{table}

%% file: manuscript/04_discussion.tex
\section{Conclusion and Discussion}
\looseness=-1
APST adapts a source-trained set-temporal decoder through unit--behavior association profiles estimated from a few labeled target-session calibration trials, with one 4-D profile interface shared by planar and high-dimensional tasks. Results on DANDI688 and FALCON support profile conditioning for cross-session decoding without target-session weight updates, while sliding-window causal attention bounds the streaming key/value state. The profile estimators and their source bases are still designed per task, and how the two conditioning sites divide the work across tasks remains to be characterized. Future work will explore unified profile estimators learned jointly with the decoder across datasets and modalities.

%% file: references.bib
@article{steinmetz2021,
  author={N. A. Steinmetz and others},
  title={Neuropixels 2.0: A miniaturized high-density probe for stable, long-term brain recordings},
  journal={Science},
  volume={372},
  number={6539},
  pages={eabf4588},
  year={2021}
}

@article{yuan2024,
  author={A. Yuan and others},
  title={Multi-day neuron tracking in high density electrophysiology recordings using {EMD}},
  journal={bioRxiv preprint 2023.08.03.551724},
  year={2024}
}

@inproceedings{jiang2024eeg,
  author={A. Jiang and S. Hou and Y. Tang and Y. Zhu},
  title={Joint spatio-temporal filtering of motion imagery {EEG} signals for data alignment in transfer learning},
  booktitle={Proc. IEEE ICASSP},
  pages={2235--2239},
  year={2024}
}

@inproceedings{rahmani2026eeg,
  author={G. Rahmani-Sane and S. Haghani},
  title={Cross-session motor imagery classification using very short {EEG} windows and unsupervised domain adaptation for real-time {BCI}},
  booktitle={Proc. IEEE ICASSP},
  year={2026}
}

@inproceedings{vaswani2017,
  author={A. Vaswani and others},
  title={Attention is all you need},
  booktitle={Proc. NeurIPS},
  volume={30},
  year={2017}
}

@inproceedings{settransformer2019,
  author={J. Lee and others},
  title={Set {Transformer}: A framework for attention-based permutation-invariant neural networks},
  booktitle={Proc. ICML},
  pages={3744--3753},
  year={2019}
}

@inproceedings{perceiver2021,
  author={A. Jaegle and others},
  title={Perceiver: General perception with iterative attention},
  booktitle={Proc. ICML},
  pages={4651--4664},
  year={2021}
}

@article{falcon2024,
  author={B. Karpowicz and others},
  title={Few-shot algorithms for consistent neural decoding ({FALCON}) benchmark},
  journal={Proc. NeurIPS},
  volume={37},
  year={2024}
}

@article{collinger2013,
  author={J. L. Collinger and others},
  title={High-performance neuroprosthetic control by an individual with tetraplegia},
  journal={The Lancet},
  volume={381},
  number={9866},
  pages={557--564},
  year={2013},
  doi={10.1016/S0140-6736(12)61816-9}
}

@inproceedings{spint2025,
  author={T. Le and others},
  title={{SPINT}: Spatial permutation-invariant neural transformer for consistent intracortical motor decoding},
  booktitle={Proc. NeurIPS},
  volume={38},
  pages={29248--29268},
  year={2025},
  doi={10.52202/085713-0870},
  url={https://proceedings.neurips.cc/paper_files/paper/2025/hash/24c8d2fe8520746f0084174b75b93ebe-Abstract-Conference.html}
}

@inproceedings{peebles2023dit,
  author={W. Peebles and S. Xie},
  title={Scalable diffusion models with transformers},
  booktitle={Proc. IEEE/CVF Int. Conf. Comput. Vis. (ICCV)},
  pages={4195--4205},
  year={2023}
}

@inproceedings{ye2023ndt2,
  author={J. Ye and J. L. Collinger and L. Wehbe and R. Gaunt},
  title={Neural Data Transformer 2: Multi-context pretraining for neural spiking activity},
  booktitle={Proc. NeurIPS},
  volume={36},
  year={2023}
}

@article{georgopoulos1982,
  author={Georgopoulos, A. P. and Kalaska, J. F. and Caminiti, R. and Massey, J. T.},
  title={On the relations between the direction of two-dimensional arm movements and cell discharge in primate motor cortex},
  journal={J. Neurosci.},
  volume={2},
  number={11},
  pages={1527--1537},
  year={1982},
  doi={10.1523/JNEUROSCI.02-11-01527.1982}
}

@article{truccolo2005,
  author={W. Truccolo and U. T. Eden and M. R. Fellows and J. P. Donoghue and E. N. Brown},
  title={A point process framework for relating neural spiking activity to spiking history, neural ensemble, and extrinsic covariate effects},
  journal={J. Neurophysiol.},
  volume={93},
  number={2},
  pages={1074--1089},
  year={2005},
  doi={10.1152/jn.00697.2004}
}

@article{hatsopoulos2007,
  author={N. G. Hatsopoulos and Q. Xu and Y. Amit},
  title={Encoding of movement fragments in the motor cortex},
  journal={J. Neurosci.},
  volume={27},
  number={19},
  pages={5105--5114},
  year={2007},
  doi={10.1523/JNEUROSCI.3570-06.2007}
}

@article{glaser2020,
  author={J. I. Glaser and others},
  title={Machine learning for neural decoding},
  journal={eNeuro},
  volume={7},
  number={4},
  year={2020}
}

@article{gallego2020,
  author={J. A. Gallego and others},
  title={Long-term stability of cortical population dynamics underlying consistent behavior},
  journal={Nat. Neurosci.},
  volume={23},
  number={2},
  pages={260--270},
  year={2020},
  doi={10.1038/s41593-019-0555-4}
}

@article{rouse2016m1,
  author={A. G. Rouse and M. H. Schieber},
  title={Spatiotemporal Distribution of Location and Object Effects in Primary Motor Cortex Neurons during Reach-to-Grasp},
  journal={J. Neurosci.},
  volume={36},
  number={41},
  pages={10640--10653},
  year={2016},
  doi={10.1523/JNEUROSCI.1716-16.2016}
}

@article{nason2021fingerbmi,
  author={S. R. Nason and others},
  title={Real-time linear prediction of simultaneous and independent movements of two finger groups using an intracortical brain-machine interface},
  journal={Neuron},
  volume={109},
  number={19},
  pages={3164--3177.e8},
  year={2021},
  doi={10.1016/j.neuron.2021.08.009}
}

@article{cunningham2014,
  author={J. P. Cunningham and B. M. Yu},
  title={Dimensionality reduction for large-scale neural recordings},
  journal={Nat. Neurosci.},
  volume={17},
  number={11},
  pages={1500--1509},
  year={2014}
}

@inproceedings{poyo2023,
  author={M. Azabou and others},
  title={A unified, scalable framework for neural population decoding},
  booktitle={Proc. NeurIPS},
  volume={36},
  pages={44937--44956},
  year={2023},
  doi={10.52202/075280-1948},
  url={https://proceedings.neurips.cc/paper_files/paper/2023/hash/8ca113d122584f12a6727341aaf58887-Abstract-Conference.html}
}

@article{degenhart2020,
  author={A. D. Degenhart and others},
  title={Stabilization of a brain--computer interface via the alignment of low-dimensional spaces of neural activity},
  journal={Nat. Biomed. Eng.},
  volume={4},
  number={7},
  pages={672--685},
  year={2020}
}

@inproceedings{farshchian2019,
  author={A. Farshchian and others},
  title={Adversarial domain adaptation for stable brain--machine interfaces},
  booktitle={Proc. ICLR},
  year={2019}
}

@article{nomad2025,
  author={B. M. Karpowicz and others},
  title={Stabilizing brain-computer interfaces through alignment of latent dynamics},
  journal={Nat. Commun.},
  volume={16},
  pages={4662},
  year={2025}
}

@article{rokni2007,
  author={U. Rokni and A. G. Richardson and E. Bizzi and H. S. Seung},
  title={Motor learning with unstable neural representations},
  journal={Neuron},
  volume={54},
  number={4},
  pages={653--666},
  year={2007}
}

@article{jarosiewicz2015,
  author={B. Jarosiewicz and others},
  title={Virtual typing by people with tetraplegia using a self-calibrating intracortical brain--computer interface},
  journal={Sci. Transl. Med.},
  volume={7},
  number={313},
  pages={313ra179},
  year={2015}
}

@misc{perich2025dandi,
  author={M. G. Perich and L. E. Miller and M. Azabou and E. L. Dyer},
  title={Long-term recordings of motor and premotor cortical spiking activity during reaching in monkeys},
  howpublished={DANDI Archive 000688, version 0.250122.1735},
  year={2025},
  doi={10.48324/dandi.000688/0.250122.1735},
  url={https://dandiarchive.org/dandiset/000688/0.250122.1735}
}

@inproceedings{perez2018film,
  author={E. Perez and F. Strub and H. de Vries and V. Dumoulin and A. Courville},
  title={{FiLM}: Visual reasoning with a general conditioning layer},
  booktitle={Proc. AAAI Conf. Artif. Intell.},
  pages={3942--3951},
  year={2018}
}

@inproceedings{press2022alibi,
  title={Train Short, Test Long: Attention with Linear Biases Enables Input Length Extrapolation},
  author={O. Press and N. A. Smith and M. Lewis},
  booktitle={Proc. ICLR},
  year={2022},
  url={https://openreview.net/forum?id=R8sQPpGCv0}
}
